\documentclass[a4paper]{article}

\usepackage{RR}
\RRNo{6215}

\usepackage{hyperref}

\usepackage{amssymb}
\usepackage{graphicx}
\usepackage{a4wide}
\usepackage{latexsym}

\usepackage{amsmath}
\usepackage{amsfonts}
\usepackage{epsfig}
\usepackage{amsmath, amsthm, amssymb}
\usepackage{hyperref}

\usepackage{verbatim}

\usepackage[dutch,english]{babel}

\usepackage{mathrsfs}

\newtheorem{Theorem}{Theorem}
\newtheorem{Proposition}[Theorem]{Proposition}

\newtheorem{Lemma}[Theorem]{Lemma}

\RRdate{June 2007}

\RRauthor{
Konstantin Avrachenkov \thanks{INRIA Sophia Antipolis, France,
e-mail: K.Avrachenkov@sophia.inria.fr}
\and Patrick Brown
\thanks{France Telecom R\&D, France, e-mail:
Patrick.Brown@orange-ftgroup.com }
\and Natalia Osipova
\thanks{INRIA Sophia Antipolis, France, e-mail:
Natalia.Osipova@sophia.inria.fr}  \thanks{The work was supported
by France Telecom R\&D Grant ``Mod\'elisation et Gestion du Trafic
R\'eseaux Internet'' no. 42937433.}}

\authorhead{K. Avrachenkov, P. Brown, N. Osipova}
\RRtitle{Le Choix du Seuil Optimal pour La File d'Attente Munie
d'une Politique à Temps Partagé Avec Deux-Niveaux}

\RRetitle{{Optimal Choice of Threshold in \\Two Level Processor
Sharing}}
\titlehead{{Optimal Choice of Threshold in Two Level Processor Sharing}}

\RRresume{
Nous étudions la file d'attente munie d'une politique a Temps Partagé
avec Deux-Niveaux "Two Level Processor Sharing" avec distribution hyper-
exponentielle des temps de service et avec le processus d'arrivée
Poisson. TLPS est un modèle commode pour ordonner l'accès aux
ressources en fonction de la taille dans un réseau TCP/IP. Dans le cas
où la distribution du temps de service est une distribution hyper-
exponentielle avec deux phases, nous trouvons une expression analytique
pour le temps de réponse moyen. Aussi nous trouvons une approximation de
valeur de seuil optimal qui réduit au minimum le temps de réponse moyen.
Dans le cas où la distribution du temps de service a plus que deux phases, nous trouvons une borne supérieure pour la fonction de temps de réponse moyen qui est conditionnée au temps de service. Nous montrons que quand la variance de la distribution des temps de service augmente, le gain dans l'exécution du système est considérable et il n'y a pas de sensibilité au choix du seuil sous optimal.
}

\RRabstract{ We analyze the Two Level Processor Sharing (TLPS)
scheduling discipline with the hyper-exponential job size
distribution and with the Poisson arrival process. TLPS is a
convenient model to study the benefit of the file size based
differentiation in TCP/IP networks. In the case of the
hyper-exponential job size distribution with two phases, we find a
closed form analytic expression for the expected sojourn time and
an approximation for the optimal value of the threshold that
minimizes the expected sojourn time. In the case of the
hyper-exponential job size distribution with more than two phases,
we derive a tight upper bound for the expected sojourn time
conditioned on the job size. We show that when the variance of the
job size distribution increases, the gain in system performance
increases and the sensitivity to the choice of the threshold near
its optimal value decreases. }

\RRmotcle{La file d'attente munie d'une politique à Temps
Partagées avec Deux-Niveaux, réseau TCP/IP, distribution
hyper-exponentielle, Laplace transforment. }

\RRkeyword{Two Level Processor sharing, Hyper-Exponential
distribution, Laplace transform.}

\RRprojet{MAESTRO}
\RRtheme{\THCom}
\URSophia 

\begin{document}

\makeRR

\section{Introduction}

$\quad$ \mbox{} It has been known for a long time that a clever
scheduling of tasks can significantly improve system performance.
For instance, Shortest Remaining Processing Time (SRPT) scheduling
discipline minimizes the expected sojourn time \cite{S68}.
However, SRPT requires to keep track of all jobs in the system and
also requires the knowledge of the remaining processing times.
These requirements are often not feasible in applications. The
examples of such applications are file size based differentiation
in TCP/IP networks \cite{RuN2C,FM03} and Web server request
differentiation \cite{GM02,HB03}.

The Two Level Processor Sharing (TLPS) scheduling discipline
\cite{kleinrock} helps to overcome the above mentioned
requirements. It uses the differentiation of jobs according to a
threshold on the attained service and gives priority to the jobs
with small sizes. A detail description of the TLPS discipline is
presented in the ensuing section. Of course, TLPS provides a
sub-optimal mechanism in comparison with SRPT. Nevertheless, as
was shown in \cite{AA06}, when the job size distribution has a
decreasing hazard rate, the performance of TLPS with appropriate
choice of threshold is very close to optimal. It turns out that
the distribution of file sizes in the Internet indeed has a
decreasing hazard rate and often could be modeled with a
heavy-tailed distributions. It is known, that the heavy-tailed
distribution could be approximated with a hyper-exponential
distribution with a significant number of phases {\cite{BM06,
feldmann_whitt}. Also in \cite{HyperExp}, it was shown that the
hyper-exponential distribution models well the file size
distribution in the Internet. Therefore, in the present work we
analyze the TLPS system with hyper-exponential job size
distribution.

The paper organization and main results are as follows. In
Section~\ref{sec:model} we provide the model formulation, main
definitions and equations. In Section~\ref{sec:twophases} we study
the TLPS discipline in the case of the hyper-exponential job size
distribution with two phases. It is known that the Internet
connections belong to two distinct classes with very different
sizes of transfer. The first class is composed of short HTTP
connections and P2P signaling connections. The second class
corresponds to downloads (PDF files, MP3 files, MPEG files, etc.).
This fact provides motivation to consider first the
hyper-exponential job size distribution with two phases.

We find an analytical expression for the expected sojourn time in
the TLPS system. Then, we present the approximation of the optimal
threshold which minimizes the expected sojourn time. We show that
the approximated value of the threshold tends to the optimal
threshold when the second moment of the job size distribution
function goes to infinity.

We show that the use of the TLPS scheduling discipline can lead to
45\% gain in the expected sojourn time in comparison with the
standard Processor Sharing. We also show that the system
performance is not too sensitive to the choice of the threshold
around its optimal value.

In Section~\ref{sec:manyphases} we analyze the TLPS discipline
when the job size distribution is hyper-exponential with many
phases. We provide an expression of the expected conditional
sojourn time as the solution of a system of linear equations. Also
we apply an iteration method to find the expression of the
expected conditional sojourn time and using the resulting
expression obtain an explicit and tight upper bound for the
expected sojourn time function. In the experimental results we
show that the relative error of the latter upper bound with
respect to the expected sojourn time function is 6-7\%.

We study the properties of the expected sojourn time function when
the parameters of the job size distribution function are selected
in a such a way that with the increasing number of phases the
variance increases. We show numerically that with the increasing
number of phases the relative error of the found upper bound
decreases. We also show that when the variance of the job size
distribution increases the gain in system performance increases
and the sensitivity of the system to the selection of the optimal
threshold value decreases.

We put some technical proofs in the Appendix.

\newpage
\section{Model description}
\label{sec:model}

\subsection{Main definitions}
\label{subsec:defin}

$\quad$ \mbox{} We study the Two Level Processor Sharing (TLPS)
scheduling discipline with the hyper-exponential job size
distribution. Let us describe the model in detail.

Jobs arrive to the system according to a Poisson process with rate
$\lambda$. We measure the job size in time units. Specifically, as
the job size we define the time which would be spent by the server
to treat the job if there were no other jobs in the system.

Let $\theta$ be a given threshold. The jobs in the system that
attained a service less than $\theta$ are assigned to the high
priority queue. If in addition there are jobs with attained service
greater than $\theta$, such a job is separated into two parts. The
first part of size $\theta$ is assigned to the high priority queue
and the second part of size $x-\theta$ waits in the lower priority
queue. The low priority queue is served when the high priority queue
is empty. Both queues are served according to the Processor Sharing
(PS) discipline.

Let us denote the job size distribution by $F(x)$. By
${\overline{F}(x)=1-F(x)}$ we denote the complementary distribution
function. The mean job size is given by $m={\int_{0}^{\infty} x
dF(x)}$ and the system load is $\rho=\lambda m$. We assume that the
system is stable ($\rho<1$) and is in steady state.

It is known that many important probability distributions
associated with network traffic are heavy-tailed. In particular,
the file size distribution in the Internet is heavy-tailed.

A distribution function has a heavy tail if \mbox{$e^{\epsilon x}
(1-F(x)) \rightarrow \infty\,\, \mbox{as}\,\, x\rightarrow \infty,
\,\,\, \forall \epsilon>0$}. The heavy-tailed distributions are
not only important and prevalent, but also difficult to analyze.
Often it is helpful to have the Laplace transform of the job size
distribution. However, there is evidently no convenient analytic
expression for the Laplace transforms of the Pareto and Weibull
distributions, the most common examples of heavy-tailed
distributions. In \cite{BM06, feldmann_whitt} it was shown that it
is possible to approximate heavy-tailed distributions by
hyper-exponential distribution with a significant number of
phases.

A hyper-exponential distribution $F_N(x)$ is a convex combination
of $N$ exponents, \mbox{$1 \leq N \leq \infty $}, namely,
\begin{eqnarray}{\label{eq:F_N}}
&& F_N(x)=1-\sum_{i=1}^N p_i e^{-\mu_i x},\quad \mu_i > 0,\,\, p_i
\geq 0,\quad i=1,...,N,\quad \mbox{and} \quad \sum_{i=1}^N{p_i}=1.
\end{eqnarray}


In particular, we can construct a sequence of hyper-exponential
distributions such that it converges to a heavy-tailed
distribution \cite{BM06}. For instance, if we select
\begin{eqnarray*}
&& p_i=\frac {\nu} {i^{\gamma_1}}, \quad
\mu_i={\frac{\eta}{i^{\gamma_2}}},\quad i=1,...,N, \\
&& \gamma_1 > 1, \quad {\frac{\gamma_1-1}{2}} < \gamma_2 <
\gamma_1-1,
\end{eqnarray*}
where ${\nu={1}/{\sum_{i=1,..,N}{{i^{-\gamma_1}}}}}, \quad
{\eta={\nu}/{m } \sum_{i=1,...,N}{{i^{\gamma_2-\gamma_1}}}},$ then
the first moment of the job size distribution is finite, but the
second moment is infinite when \mbox{$N \rightarrow \infty$}.
Namely, the first and the second moments $m$ and $d$ for the
hyper-exponential distribution are given by:
\begin{eqnarray}{\label{eq:m_d}}
&& m = \int_0^{\infty}{x\, dF(x)}=
\sum_{i=1}^N{{\frac{p_i}{\mu_i}}}, \quad  d
={\int_0^{\infty}{x^2\, dF(x)}}=2\sum_{i=1}^N
{\frac{p_i}{\mu_i^2}}.
\end{eqnarray}
Let us denote
\begin{eqnarray}{\label{eq:F_i_theta}}
&& \overline{F_{\theta}^i}=p_i e^{-\mu_i \theta }, \quad
i=1,...,N.
\end{eqnarray}
We note that \mbox{$\sum_{i} {\overline{F_{\theta}^i}}
=\overline{F}( \theta )$}. The hyper-exponential distribution has
a simple Laplace transform:
\begin{eqnarray*}
&& L_{\overline{F}(x)}(s) = \sum_{i=1}^N{\frac{p_i
\mu_i}{s+\mu_i}}.
\end{eqnarray*}

We would like to note that the hyper-exponential distribution has
a decreasing hazard rate. In {\cite{AA06}} it was shown, that when
a job size distribution has a decreasing hazard rate, then with
the selection of the threshold the expected sojourn time of the
TLPS system could be reduced in comparison to standard PS system.


Thus, in our work we use hyper-exponential distributions to
represent job size distribution functions. In particular, the
application of the hyper-exponential job size distribution with
two phases is motivated by the fact that in the Internet
connections belong to two distinct classes with very different
sizes of transfer. The first class is composed of short HTTP
connections and P2P signaling connections. The second class
corresponds to downloads (PDF files, MP3 files, MPEG files, etc.).
So, in the first part of our paper we look at the case of the
hyper-exponential job size distribution with two phases and in the
second part of the paper we study the case of more than two
phases.

\subsection{The expected sojourn time in TLPS system}{\label{subsec:exp_soj_time_TLPS}}

$\quad$ \mbox{} Let us denote by $\overline{T}^{TLPS}(x)$ the
expected conditional sojourn time in the TLPS system for a job of
size $x$. Of course, $\overline{T}^{TLPS}(x)$ also depends on
$\theta$, but for expected conditional sojourn time we only
emphasize the dependence on the job size. On the other hand, we
denote by $\overline{T}(\theta)$ the overall expected sojourn time
in the TLPS system. Here we emphasize the dependence on $\theta$
as later we shall optimize the overall expected sojourn time with
respect to the threshold value.

To calculate the expected sojourn time in the TLPS system we need
to calculate the time spent by a job of size $x$ in the first high
priority queue and in the second low priority queue. For the jobs
with size \mbox{$x \leq \theta$} the system will behave as the
standard PS system where the service time distribution is
truncated at $\theta$. Let us denote by
\begin{eqnarray} {\label{eq:Xn}}
&& {\overline{X_{\theta}^n}} = \int_{0}^{ \theta }\!n y^{n-1}
\overline{F}(y) dy
\end{eqnarray}
the $n$-th moment of the distribution truncated at $\theta$. In
the following sections we will need
\begin{eqnarray}{\label{eq:X1X2}}
&& \overline{X_{\theta}^1}=
m-{\sum_{i=1}^N{\frac{\overline{F_{\theta}^i}}{\mu_i}}}, \quad
\overline{X_{\theta}^2}=2\sum_{i=1}^N{\frac{p_i}{\mu_i^2}} -2
\theta \left(m- \sum_{i=1}^N
{\frac{\overline{F_{\theta}^i}}{\mu_i}} \right)- 2{\sum_{i=1}^N
{\frac{\overline{F_{\theta}^i}} {\mu_i^2}}}.
\end{eqnarray}

The utilization factor for the truncated distribution is
\begin{eqnarray}{\label{eq:rho_theta}}
&& {\rho_{ \theta }= \lambda\,{  \overline{X_{ \theta }^1}}}=\rho-
\lambda {\sum_{i=1}^N{\frac{\overline{F_{\theta}^i}}{\mu_i}}}.
\end{eqnarray}

Then, the expected conditional response time is given by
\begin{displaymath}
\overline{T}^{TLPS}(x)= \left\{
\begin{array}{l l}
{\displaystyle \frac{x}{1-\rho_{ \theta }}}, & x\in[0, \theta ], \\
{\displaystyle \frac {\overline{W}( \theta )+ \theta +\alpha(x-
\theta )}{1-\rho_{ \theta }}}, & x\in(\theta,\infty).
\end{array}\right.
\end{displaymath}

According to {\cite{kleinrock}}, here \mbox{${(\overline{W}(
\theta )+ \theta)/{(1-\rho_{ \theta })}}$} expresses the time
needed to reach the low priority queue. This time consists of the
time \mbox{${\theta/{(1-\rho_{ \theta })}}$} spent in the high
priority queue, where the flow is served up to the threshold
$\theta$, plus the time \mbox{${\overline{W}( \theta )/{(1-\rho_{
\theta })}}$} which is spent waiting for the high priority queue
to empty. Here \mbox{$\overline{W}(\theta )= {\lambda{
\overline{X_{ \theta }^2}}}/ {(2(1-\rho_{ \theta }))}$}.

The remaining term \mbox{${{\alpha(x- \theta )}/{(1-\rho_{ \theta
})}}$} is the time spent in the low priority queue. To find
$\alpha(x)$ we can use the interpretation of the lower priority
queue as a PS system with batch arrivals {\cite{AAB_BPS_MLPS_QS,
Osipova_rap-rech}}. As was shown in \cite{kleinrock},
\mbox{$\alpha'(x)=d\alpha/dx$} is the solution of the following
integral equation
\begin{equation}
\label{IntAlpha} {\alpha'(x)=\lambda \overline{n}
\int_{0}^{\infty} \alpha'(y) \overline{B}(x+y)dy + \lambda
\overline{n} \int_{0}^{x} \alpha'(y) \overline{B}(x-y)dy +b
\overline{B}(x)+1}.
\end{equation}
Here 
\mbox{$\overline{n}=\overline{F}(\theta)/(1-\rho_{\theta})$} is
the average batch size,  
\mbox{${\overline{B}(x)}= \overline{F}(\theta +
x)/ \overline{F}(\theta)$} is the complementary truncated
distribution and
\mbox{$b=b(\theta)=2 \lambda \overline{F}(\theta)
(\overline{W}(\theta) + \theta)/(1-\rho_{\theta})$} is the average
number of jobs that arrive to the low priority queue in addition
to the tagged job.

The expected sojourn time in the system is given by the following
equations:
\begin{eqnarray}
&& {\overline{T}(\theta)}= \int_0^{\infty} {\overline T}^ {TLPS}
(x) dF(x),\nonumber\\
&& \overline{T}(\theta)=
\frac{{\overline{X_{\theta}^1}}+\overline{W}( \theta
)\overline{F}( \theta )}{1-\rho_{ \theta }} +
{\frac{1}{1-\rho_{\theta}}}\overline{T}^{BPS}(\theta),{\label{eq:T_T_BPS}}\\
&& \overline{T}^{BPS}(\theta) = {\int_{ \theta}^{\infty}\!\!
{{\alpha(x- \theta )}}dF(x)=\int_0^\infty\!\!
\alpha'(x)\overline{F}(x+\theta)dx}.{\label{eq:T_BPS}}
\end{eqnarray}

\newpage
\section{Hyper-exponential job size distribution with two phases}
\label{sec:twophases}

\subsection{Notations}

$\quad$ \mbox{} In the first part of our work we consider the
hyper-exponential job size distribution with two phases. Namely,
according to ({\ref{eq:F_N}}) the cumulative distribution function
$F(x)$ for $N=2$ is given by
\begin{eqnarray*}
&& F(x) = 1-p_1 {e^{-\mu_1 x}}-p_2 {e^{-\mu_2 x}},
\end{eqnarray*}
where \mbox{$p_1+p_2=1$} and \mbox{$p_1, p_2>0$}.

The mean job size $m$, the second moment $d$, the parameters
$\overline{F_{\theta}^i}$, $\overline{X_{\theta }^1}$,
$\overline{X_{ \theta }^2}$ and $\rho_{\theta}$ are defined as in
Section~{\ref{subsec:defin}} and
Section~{\ref{subsec:exp_soj_time_TLPS}} by formulas
({\ref{eq:m_d}}),({\ref{eq:F_i_theta}}),({\ref{eq:X1X2}}),
({\ref{eq:rho_theta}}) with $N=2$.

We note that the system has four free parameters. In particular,
if we fix $\mu_1$, \mbox{$\epsilon=\mu_2/\mu_1$}, $m$, and $\rho$,
the other parameters $\mu_2$, $p_1$, $p_2$ and $\lambda$ will be
functions of the former parameters.

\subsection{Explicit form for the expected sojourn time}

$\quad$ \mbox{} To find $\overline{T}^{TLPS}(x)$ we need to solve
the integral equation (\ref{IntAlpha}). To solve (\ref{IntAlpha})
we use the Laplace transform based method described in
\cite{bansal}.

\begin{Theorem}{\label{teor:T_solve_two_phase}}
The expected sojourn time in the TLPS system with the
hyper-exponential job size distribution with two phases is given
by
\begin{equation}{\label{eq:T_solve_2_phase}}
{\overline{T}(\theta)}= \frac{\overline{X_{\theta}^1}
+\overline{W}(\theta)\overline{F}( \theta )}{1-\rho_{ \theta }}+
{\frac{m-\overline{X_{ \theta }^1}}{1-\rho}}+{\frac {b(\theta)
\left({\mu_1 \mu_2}(m-\overline{X_{\theta}^1})^2
+{\delta_{\rho}(\theta)}{\overline{F}\,^2(\theta)} \right)}
{2(1-\rho){\overline{F}(\theta)}{\left(\mu_1+\mu_2-{\gamma(\theta)
\overline{F}(\theta)}\right)}}},
\end{equation} where
$ \delta_{\rho}(\theta )=1-{\gamma(\theta)}(m- \overline{X_{
\theta }^1})=(1-\rho)/(1-\rho_{\theta})$ and  $\gamma( \theta )=
{\lambda}/{(1-\rho_{ \theta })}$.
\end{Theorem}
\begin{proof}[\textsc{\textbf{Proof.}}]
We can rewrite integral equation (\ref{IntAlpha}) in the following
way
\begin{eqnarray*}
\alpha'(x)&=&\gamma( \theta ) \int_{0}^{\infty} \!\!\alpha'(y)
{{\overline{F}(x+y+\theta )}}dy + \gamma( \theta )\int_{0}^{x}\!\!
\alpha'(y){\overline{F}( x-y+\theta )}dy + b(\theta) \overline{B}(x)+1, \\
\alpha'(x) &=& \gamma( \theta ) \sum_{i=1,2}{
\overline{F_{\theta}^i}} e^{{-\mu_i x}} \int_{0}^{\infty}\!\!
\alpha'(y) e^{-\mu_i y}dy +\gamma( \theta ) \int_{0}^{x}\!\!
\alpha'(y){\overline{F}(x-y+\theta)}dy +b(\theta)
\overline{B}(x)+1.
\end{eqnarray*}
We note that in the latter equation \mbox{$\int_{0}^{\infty}
\alpha'(y){e^{-\mu_i y}} dy$, $i=1,2$} are the Laplace transforms
of $\alpha'(y)$ evaluated at \mbox{$\mu_i, i=1,2$}. Denote
\begin{eqnarray*}
&& L_i=\int_{0}^{\infty} \alpha'(y){e^{-\mu_i y} dy },\qquad
i=1,2.
\end{eqnarray*} Then, we have
\begin{eqnarray*}
&& \alpha'(x) =\gamma(\theta) \sum_{i=1,2}
{\overline{F_{\theta}^i}} L_i e^{{-\mu_i x}} +\gamma(\theta)
\int_{0}^{x} \alpha'(y) {\overline{F}( x-y+\theta)}dy +b(\theta)
\overline{B}(x)+1.
\end{eqnarray*}
Now taking the Laplace transform of the above equation and using
the convolution property, we get
\begin{eqnarray*}
&& L_{\alpha'}(s) = \gamma(\theta) \sum_{i=1,2}{\frac{
\overline{F_{\theta}^i}L_i}{s+\mu_i}}+\gamma(\theta)
\sum_{i=1,2}{\frac{\overline{F_{\theta}^i}
L_{\alpha'}(s)}{s+\mu_i}}+ \frac{b(\theta)}{\overline{F}(\theta)}
\sum_{i=1,2}{\frac{ \overline{F_{\theta}^i}}{s+\mu_i}} + {\frac 1
s},\\
\Longrightarrow && L_{\alpha'}(s) \left (1- \gamma(\theta)
\sum_{i=1,2}{\frac{\overline{F_{\theta}^i}}{s+\mu_i}}\right)=
\gamma(\theta) \sum_{i=1,2}{\frac{\overline{F_{\theta}^i}
L_i}{s+\mu_i}} + \frac{b(\theta)}{\overline{F}(\theta)}
\sum_{i=1,2}{\frac {\overline{F_{\theta}^i}}{s+\mu_i}} + {\frac 1
s}.
\end{eqnarray*}
Here \mbox{$L_{\alpha'}(s)
=\int_{0}^{\infty}{\alpha'(x)e^{-sx}dx}$} is the Laplace transform
of $\alpha'(x)$. Let us note that\\
\mbox{$L_{\alpha'}(\mu_i)=L_i,\,\,\, i=1,2$}. Then, if we
substitute into the above equation \mbox{$s=\mu_1$} and
\mbox{$s=\mu_2$}, we can get $L_1$ and $L_2$ as a solution of the
linear system
\begin{eqnarray*}
L_1 &=& {\frac {1} {\left(\mu_1+\mu_2-{\gamma(\theta)
\overline{F}(\theta)}\right){\delta_{\rho}(\theta)}}}\left( {\frac
{b( \theta )} {2 \overline{F}( \theta )}} \left(
{\mu_2}(m-\overline{X_{ \theta }^1})+{\delta_{\rho}(\theta)}{
\overline{F}( \theta )} \right)\right) + {{\frac {1} {\mu_1
{\delta_{\rho}(\theta)}}}},\\
L_2 &=&{\frac {1} {\left(\mu_1+\mu_2-{\gamma(\theta)
\overline{F}(\theta)}\right){\delta_{\rho}(\theta)}}}\left( {\frac
{b( \theta )} {2{\overline{F}( \theta )}}} \left( {{\mu_1}
(m-\overline{X_{ \theta
}^1})}+{\delta_{\rho}(\theta)}{\overline{F}( \theta )}
\right)\right) + {\frac {1}{\mu_2 \delta_{\rho}(\theta)}}.
\end{eqnarray*}
Next we need to calculate \mbox{$\overline{T}^{BPS}(\theta)$}.
\begin{eqnarray*}
\overline{T}^{BPS}(\theta) &=&
\int_{0}^{\infty}{{\alpha'(x)}}\overline{F}(x+ \theta )dx=
\int_{0}^{\infty}{{\alpha'(x)}}\sum_{i=1,2}{\overline{F_{\theta}^i}
e^{-\mu_i x}}dx =\sum_{i=1,2}{{\overline{F_{\theta}^i} L_i}},\\
\overline{T}^{BPS}(\theta)&=&{\frac{1-\rho_{\theta}}{1-\rho}}
\left({{m-\overline{X_{ \theta }^1}}}+{\frac{b(\theta) \left(
{\mu_1 \mu_2}(m-\overline{X_{ \theta}^1})^2
+{\delta_{\rho}(\theta)}{\overline{F}\,^2(\theta)}\right)}
{2{\overline{F}( \theta )}{\left(\mu_1+\mu_2-{\gamma(\theta)
\overline{F}(\theta)}\right)}}} \right).
\end{eqnarray*}
Finally, by (\ref{eq:T_T_BPS}) we have
({\ref{eq:T_solve_2_phase}}). \end{proof}

\subsection{Optimal threshold approximation}

$\quad$ \mbox{} We are interested in the minimization of the
expected sojourn time ${\overline {T}(\theta)}$ with respect to
$\theta$. Of course, one can differentiate the exact analytic
expression provided in Theorem~{\ref{teor:T_solve_two_phase}} and
set the result of the differentiation to zero. However, this will
give a transcendental equation for the optimal value of the
threshold.

In order to find an approximate solution of
\mbox{${\overline{T}'(\theta)}=d\overline{T}(\theta)/d\theta =0$},
we shall approximate the derivative ${\overline{T}'(\theta)}$ by
some function ${\widetilde{\overline{T}}}\!\,'{(\theta)}$ and
obtain a solution to \mbox{$\widetilde{\overline{T}}\!\,'
{(\tilde\theta_{opt})}=0$}.

Since in the Internet connections belong to two distinct classes
with very different sizes of transfer (see Section
{\ref{subsec:defin}}), then to find the approximation of
${\overline{T}'(\theta)}$ we consider a particular case when
\mbox{$\mu_2 << \mu_1$}. Let us introduce a small parameter
$\epsilon$ such that
\begin{eqnarray*}
&& \mu_2=\epsilon {\mu_1}, \quad \epsilon \rightarrow 0,\quad
 p_1=1-{\frac {\epsilon\, \left( m{  \mu_1}-1 \right)
}{1-\epsilon}} ,\quad p_2={\frac {\epsilon\, \left( m{ \mu_1}-1
\right) }{1-\epsilon}}.
\end{eqnarray*}
We note that when \mbox{$\epsilon \to 0$} the second moment of the
job size distribution goes to infinity.

We then verify that $\tilde\theta_{opt}$ indeed converges to the
minimum of $\overline{T}(\theta)$ when \mbox{$\epsilon \rightarrow
0$}.

\begin{Lemma}
\label{lemma:lambda1rho} The following inequality holds:
\mbox{$\mu_1 \rho > \lambda$}.
\end{Lemma}
\begin{proof}[\textsc{\textbf{Proof.}}]
Since $p_1>0$ and $p_2>0$, we have the following inequality
$m\mu_1>1$. Then, \mbox{$m>\frac{1}{\mu_1}$}. Taking into account
that \mbox{$\lambda m = \rho$} we get \mbox{$\frac{\rho}{\lambda}
> \frac{1}{\mu_1}$}. Consequently, we have that \mbox{$\mu_1 \rho >
\lambda$}.\end{proof}

%
%

\begin{Proposition}{\label{prop:T_diff_approx}}
The derivative of $\overline{T}(\theta)$ can be approximated by the
following function:
\begin{eqnarray*}
&& {{\widetilde{\overline{T}}\!\,'}(\theta)}=-e^{-\mu_1 {\theta}}
{\mu_1}{c_1}+ e^{-\mu_2 {\theta}} {\mu_2}{c_2},
\end{eqnarray*} where
\begin{eqnarray} {\label{eq:c_1}}
c_1={\frac{\lambda(m \mu_1 - 1)}
{\mu_1(\mu_1-\lambda)(1-\rho)}}, \quad c_2 = \frac{\lambda (m
\mu_1-1)} {(\mu_1-\lambda)^2}.
\end{eqnarray}
Namely,
\begin{eqnarray*}
&& {\overline{T}'(\theta)}-
{{\widetilde{\overline{T}}\!\,'}(\theta)} =O(\mu_2/\mu_1).
\end{eqnarray*}
\end{Proposition}
\begin{proof}[\textsc{\textbf{Proof.}}]
Using the analytical expression for both ${\overline{T}'(\theta)}$
and ${{\widetilde{\overline{T}}\!\,'}(\theta)}$, we get the Taylor
series for \mbox{${\overline{T}'(\theta)}-
{{\widetilde{\overline{T}}\!\,'}(\theta)}$} with respect to
$\epsilon$, which shows that indeed
\begin{eqnarray*}
&& \overline{T}'(\theta)- {\widetilde{\overline{T}}\!\,'}(\theta)
= O(\epsilon). \end{eqnarray*} \end{proof}

Thus we have found an approximation of the derivative of
$\overline{T}(\theta)$. Now we can find an approximation of the
optimal threshold by solving the equation
\mbox{${\widetilde{\overline{T}}\!\,'}(\theta)=0$}.

\begin{Theorem}
Let $\theta_{opt}$ denote the optimal value of the threshold.
Namely, \mbox{$\theta_{opt}=\mbox{\rm arg} \min
\overline{T}(\theta)$}. The value $\tilde\theta_{opt}$ given by
\begin{eqnarray*}
&&\tilde\theta_{opt}=\frac{1}{\mu_1-\mu_2} \ln\left(
\frac{(\mu_1-\lambda)}{\mu_2(1-\rho)}\right)
\end{eqnarray*}
approximates $\theta_{opt}$ so that
\mbox{$\overline{T}'(\tilde\theta_{opt})=\mbox{\rm
o}(\mu_2/\mu_1)$}.
\end{Theorem}
\begin{proof}[\textsc{\textbf{Proof.}}]
Solving the equation
\begin{eqnarray*}
&& {{\widetilde{\overline{T}}\!\,'}(\theta)}=0,
\end{eqnarray*} we get an analytic expression for the approximation of the optimal
threshold:
\begin{eqnarray*}
&& {\widetilde{\theta}_{opt}=-{\frac{1}{\mu_1 (1-\epsilon)}} \ln
\left(\epsilon \frac {\mu_1{(1-\rho)}} {(\mu_1-\lambda)}\right)
={\frac{1}{\mu_1-\mu_2}} \ln \left(\frac
{(\mu_1-\lambda)}{\mu_2{(1-\rho)}}\right) }.
\end{eqnarray*}

Let us show that the above threshold approximation is greater than
zero. We have to show that \mbox{$\frac
{(\mu_1-\lambda)}{\mu_2{(1-\rho)}}>1$}. Since \mbox{$\mu_1
> \mu_2$} and \mbox{$\mu_1 \rho > \lambda$} (see Lemma~\ref{lemma:lambda1rho}), we have
\begin{eqnarray*}
&& \mu_1 >{\mu_2} \\
\Longrightarrow &&\mu_1 (1-\rho)>{\mu_2{(1-\rho)}} \\
\Longrightarrow &&{\lambda}<\mu_1 \rho<{\mu_1
-\mu_2{(1-\rho)}} \\
\Longrightarrow && {(\mu_1-\lambda)}>{\mu_2{(1-\rho)}}.
\end{eqnarray*}

Expanding ${{\overline{T}}'(\widetilde{\theta}_{opt})}$ as a power
series with respect to $\epsilon$ gives:
\begin{eqnarray*}
&& {\overline{T}'(\widetilde{\theta}_{opt})}=\epsilon^2 (const_0 +
const_1 \ln{\epsilon}+ const_2 \ln^2{\epsilon}),
\end{eqnarray*}
where \mbox{$const_i,\,\, i=1,2$} are some constant
values{\footnote{The expressions for the constants $const_i$ are
cumbersome and can be found using Maple command ``series''.}} with
respect to $\epsilon$. Thus,
\begin{eqnarray*}
&& {{\overline{T}}'(\widetilde{\theta}_{opt})}=
o(\epsilon)=o(\mu_2/\mu_1),
\end{eqnarray*} which completes the proof.
\end{proof}

In the next proposition we characterize the limiting behavior of
$\overline{T}({\theta}_{opt})$ and
$\overline{T}(\widetilde{\theta}_{opt})$ as \mbox{$\epsilon \to
0$}. In particular, we show that
$\overline{T}(\widetilde{\theta}_{opt})$ tends to the exact
minimum of $\overline{T}(\theta)$ when \mbox{$\epsilon \to 0$}.
\begin{Proposition}
\begin{eqnarray*}
&&{\lim_{\epsilon \rightarrow 0}{\overline{T}({\theta}_{opt})}=
\lim_{\epsilon \rightarrow
0}{\overline{T}(\widetilde{\theta}_{opt})}}
={\frac{m}{1-\rho}}-c_1,
\end{eqnarray*} where $c_1$ is given by ({\ref{eq:c_1}}).
\end{Proposition}
\begin{proof}[\textsc{\textbf{Proof.}}]
We find the following limit, when \mbox{$\epsilon \to 0$}:
\begin{eqnarray*}
&& \lim_{\epsilon \rightarrow
0}{\overline{T}({\theta})}={\frac{m}{1-\rho}}-{\frac{\lambda( m
\mu_1-1)} {\mu_1(\mu_1 -\lambda)(1-\rho)}}+\frac{\lambda(
 m \mu_1-1) e^{-\mu_1 \theta}}{ \mu_1(\mu_1 -\lambda)(1-\rho)},\\
&& {\lim_{\epsilon \rightarrow
0}{\overline{T}({\theta})}={\frac{m}{1-\rho}}-c_1+c_1 e^{-\mu_1
\theta}},
\end{eqnarray*}
where $c_1$ is given by ({\ref{eq:c_1}}). Since the function
\mbox{${\lim_{\epsilon \rightarrow 0}{\overline{T}({\theta})}}$}
is a decreasing function, the optimal threshold for it is
\mbox{$\theta_{opt}=\infty$}. Thus,
\begin{eqnarray*}
&& {\lim_{\epsilon \rightarrow 0} {\overline{T}({\theta_{opt}})}}=
{\lim_{\theta \rightarrow \infty}\lim_{\epsilon \rightarrow
0}{\overline{T}({\theta})} ={\frac{m}{1-\rho}}-c_1}.
\end{eqnarray*}
On the other hand, we obtain
\begin{eqnarray*}
&& {\lim_{\epsilon \rightarrow 0}
{\overline{T}({\widetilde{\theta}_{opt}})}
={\frac{m}{1-\rho}}-c_1},
\end{eqnarray*}
which proves the proposition.
\end{proof}

\subsection{Experimental results}

In Figure~1-2 we show the plots for the following parameters:
\mbox{$\rho=10/11$} (default value), \mbox{$m=20/11$},
\mbox{$\mu_1=1$}, \mbox{$\mu_2=1/10$}, so \mbox{$\lambda=1/2$} and
\mbox{$\epsilon=\mu_2/\mu_1=1/10$}. Then, \mbox{$p_1=10/11$} and
\mbox{$p_2=1/11$}.

In Figure~1 we plot ${{\overline{T}}({\theta})}$,
${\overline{T}^{PS}}$ and
${{\overline{T}}(\widetilde{\theta}_{opt})}$. We note, that the
expected sojourn time in the standard PS system
$\overline{T}^{PS}$ is equal to ${{\overline{T}}(0)}$. We observe
that ${{\overline{T}}(\widetilde{\theta}_{opt})}$ corresponds well
to the optimum even though \mbox{$\epsilon=1/10$} is not too
small.

Let us now study the gain that we obtain using TLPS, by setting
\mbox{$\theta = \widetilde{\theta}_{opt}$}, in comparison with the
standard PS. To this end, we plot the ratio \mbox{${g
(\rho)={\frac {{\overline{T}^{PS}}-{\overline{T}}
(\widetilde{\theta}_{opt})} {{\overline{T}}^{PS}}}}$} in Figure~2.
The gain in the system performance with TLPS in comparison with PS
strongly depends on $\rho$, the load of the system. One can see,
that the gain of the TLPS system with respect to the standard PS
system goes up to 45\% when the load of the system increases.

To study the sensitivity of the TLPS system with respect to
$\theta$, we find the gain of the TLPS system with respect to the
standard PS system, we plot in Figure~2 \mbox{${g_1(\rho)={\frac
{{\overline{T}^{PS}}-{\overline{T}}({\frac{3}{2}}\widetilde{\theta}_{opt})}
{{\overline{T}}^{PS}}}}$} and \mbox{${g_2 (\rho)={\frac
{{\overline{T}^{PS}}-{\overline{T}}({\frac{1}{2}}\widetilde{\theta}_{opt})}
{{\overline{T}}^{PS}}}}$}. Thus, even with the 50\% error of the
$\widetilde{\theta}_{opt}$ value, the system performance is close
to optimal.

One can see that it is beneficial to use TLPS instead of PS in the
case of heavy and moderately heavy loads. We also observe that the
optimal TLPS system is not too sensitive to the choice of the
threshold near its optimal value, when the job size distribution
is hyper-exponential with two phases. Nevertheless, it is better
to choose larger rather than smaller values of the threshold.

 \begin{figure}[t]
   \begin{minipage}{0.48\linewidth}
        \centering {\epsfxsize=3.2 in \epsfbox{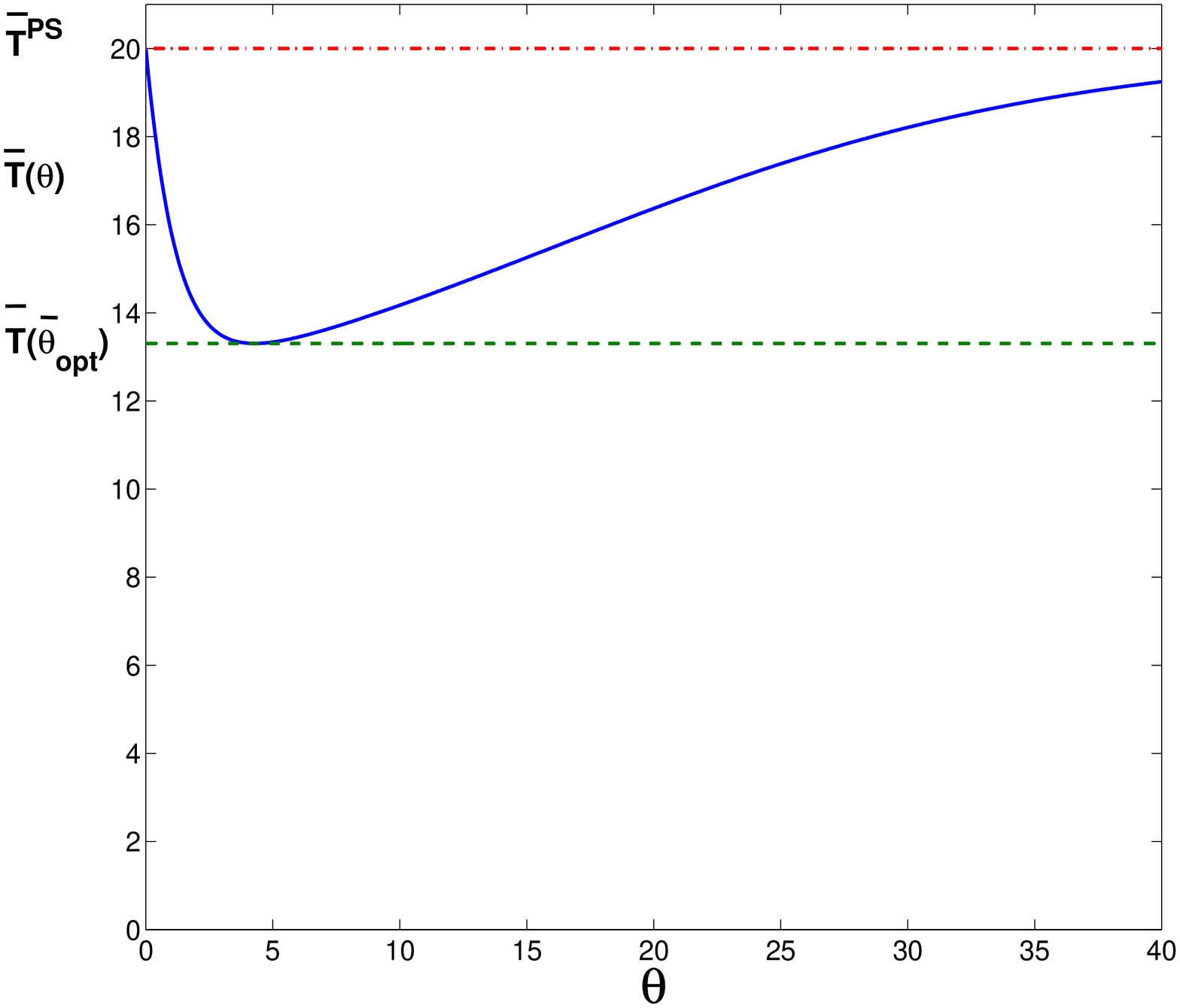}}
        \caption{$\overline{T}(\theta)$ - solid line, $\overline{T}^{PS}(\theta)$ - dash dot line,
$\overline{T}(\widetilde{\theta}_{opt})$ - dash line}
   \end{minipage}
   \hfill
   \begin{minipage}{0.48\linewidth}
        \centering {\epsfxsize=3.2 in \epsfbox{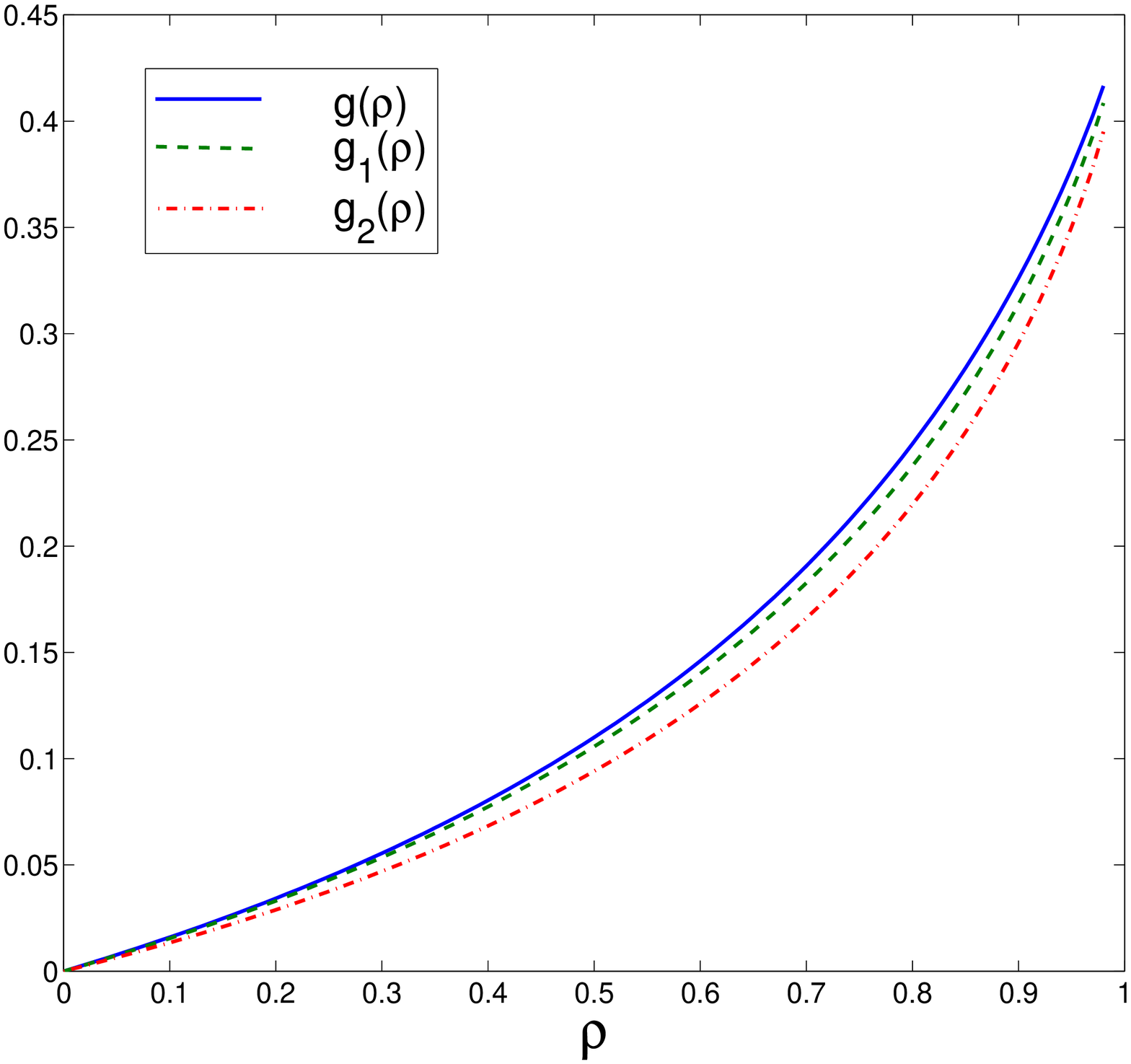}}
        \caption {$g(\rho)$ - solid line, $g_1(\rho)$ - dash line, $g_2(\rho)$ - dash dot line}
   \end{minipage}
 \end{figure}

\newpage
\section{Hyper-exponential job size distribution with more than two phases}
\label{sec:manyphases}

\subsection{Notations}

$\quad$ \mbox{} In the second part of the presented work we
analyze the TLPS discipline with the hyper-exponential job size
distribution with more than two phases. As was shown in
\cite{BM06,HyperExp,feldmann_whitt}, the hyper-exponential
distribution with a significant number of phases models well the
file size distribution in the Internet. Thus, in this section as
the job size distribution  we take the hyper-exponential function
with many phases. Namely, according to ({\ref{eq:F_N}}),
\begin{eqnarray*}
&& F(x)=1-\sum_{i=1}^{N}{p_i \, e^{-\mu_i \, x}}, \quad
{\sum_{i=1}^{N}{p_i}=1}, \,\,\, \mu_i > 0, \,\,p_i \geq 0,\,\,\,
i=1,...,N, \,\,\, 1<N \leq \infty.
\end{eqnarray*}
In the following we shall write simply $\sum_{i}$ instead of
$\sum_{i=1}^{N}$.
\\
The mean job size $m$, the second moment $d$, the parameters
$\overline{F_{\theta}^i}$, $\overline{X_{\theta }^1}$,
$\overline{X_{ \theta }^2}$ and $\rho_{\theta}$ are defined as in
Section~{\ref{subsec:defin}} and
Section~{\ref{subsec:exp_soj_time_TLPS}} by formulas
({\ref{eq:m_d}}),({\ref{eq:F_i_theta}}),({\ref{eq:X1X2}}),
({\ref{eq:rho_theta}}) for any \mbox{$1 \leq N \leq \infty$}.
The formulas presented in Section~\ref{subsec:exp_soj_time_TLPS}
can still be used to calculate $b(\theta)$, $\overline{B}(x)$,
$\overline{W}(\theta)$, $\gamma(\theta)$, $\delta_{\rho}(\theta)$,
$\overline{T}^{TLPS}(x)$, $\overline{T}(\theta)$.

We shall also need the following operator notations:
\begin{eqnarray}
{\Phi_{1}(\beta(x))}&=&{\gamma(\theta)\int_{0}^{\infty} \beta(y)
{{\overline{F}( x+y+\theta)}}dy} +{\gamma(\theta){ \int_{0}^{x}
\beta(y){\overline{F}( x-y+\theta)}dy }}, \\
{\Phi_2(\beta(x))}&=&{\int_{0}^{\infty} \beta(y) {{\overline{F}(
y+\theta)}}dy}
\end{eqnarray}
for any function $\beta(x)$. In particular, for some given
constant c, we have
\begin{eqnarray}
{\Phi_1(c)}&=&{c\, \gamma(\theta)(m-{\overline{X_{ \theta}^1}})=c\, q},\label{eq:Phi_1_const}\\
{\Phi_2(c)}&=&{c\, (m-{\overline{X_{\theta}^1}})},
\label{eq:Phi_2_const}
\end{eqnarray}
where
\begin{equation}
\label{q_val} {q}={\gamma(\theta)(m-{\overline{X_{
\theta}^1}})=\frac{\lambda(m-{\overline{X_{
\theta}^1}})}{1-\rho_{\theta}}=
\frac{\rho-\rho_{\theta}}{1-\rho_{\theta}}<1}.
\end{equation}
The integral equation (\ref{IntAlpha}) can now be rewritten in the
form
\begin{equation}{\label{eq:alphadiff_Phi}}
{\alpha'(x) = \Phi_1(\alpha'(y))}{
+\frac{b(\theta)}{\overline{F}(\theta)} \overline{F}(x+\theta)+1}.
\end{equation} and equation
(\ref{eq:T_BPS}) for $\overline{T}^{BPS}(\theta)$ takes the form
\begin{equation}{\label{eq:T_BPS_Phi}}
{\overline{T}^{BPS}(\theta)=\Phi_2(\alpha'(x))}.
\end{equation}

\subsection{Linear system based solution}

$\quad$ \mbox{} Similarly to the first part of the proof of
Theorem~\ref{teor:T_solve_two_phase}, we obtain the following
proposition.

\begin{Proposition}{\label{propos:linear_sys_solut}}
\begin{eqnarray*}
\overline{T}^{BPS}(\theta)=\sum_i \overline{F_{\theta}^i}L_i,
\end{eqnarray*} with
\begin{eqnarray*}
{L_i=L^{*}_i+\frac{1}{\delta_{\rho}(\theta) \mu_i}},
\end{eqnarray*}
where the $L^{*}_i$ are the solution of the linear system
\begin{equation}{\label{Li_part1}}
L^{*}_p \left (1- \gamma( \theta ) \sum_{i}
{\frac{\overline{F_{\theta}^i}} {\lambda_p+\mu_i}}\right)=
\gamma(\theta) \sum_{i} {\frac {{ \overline{F_{\theta}^i}}
L^{*}_i}{\lambda_p+\mu_i}}+{\frac{b(\theta)}{\overline{F}( \theta
)} \sum_{i}{\frac{ \overline{F_{\theta}^i}}
{\lambda_p+\mu_i}}},\quad p=1,...,N.
\end{equation}
\end{Proposition}

Unfortunately, the system (\ref{Li_part1}) does not seem to have a
tractable finite form analytic solution. Therefore, in the ensuing
subsections we proposed an alternative solution based on an
operator series and construct a tight upper bound.

\subsection{Operator series form for the expected sojourn time}

$\quad$ \mbox{} Since the operator $\Phi_1$ is a contraction
\cite{RuN2C,AAB_BPS_MLPS_QS}, we can iterate equation
(\ref{eq:alphadiff_Phi}) starting from some initial point
$\alpha'_0$. The initial point could be simply a constant. As
shown in \cite{RuN2C,AAB_BPS_MLPS_QS} the iterations will converge
to the unique solution of (\ref{eq:alphadiff_Phi}). Specifically,
we make iterations in the following way:
\begin{equation}{\label{alpha_n_1}} {\alpha'_{n+1}(x) = \Phi_1(\alpha'_{n}(x))}{
+\frac{b(\theta)}{\overline{F}(\theta)} \overline{F}(x+\theta)+1},
\quad n=0,1,2,...\end{equation} At every iteration step we
construct the following approximation of
$\overline{T}^{BPS}(\theta)$ according to ({\ref{eq:T_BPS_Phi}}):
\begin{equation}{\label{T_BPS_n_1}} {\overline{T}^{BPS}_{n+1}(\theta)=\Phi_2(\alpha'_{n+1}(x))}.\end{equation}
%

Using (\ref{alpha_n_1}) and (\ref{T_BPS_n_1}), we can construct
the operator series expression for the expected sojourn time in
the TLPS system.

\begin{Theorem}
The expected sojourn time $\overline{T}(\theta)$ in the TLPS
system with the hyper-exponential job size distribution is given
by
\begin{eqnarray}\label{eq:T_oper}
&& {\overline{T}(\theta)} = \frac{{\overline{X_{
\theta}^1}}+\overline{W}( \theta )\overline{F}( \theta )}{1-\rho_{
\theta }}+{\frac{m-{\overline{X_{ \theta}^1}}}{1-\rho}} +
{\frac{b(\theta)}{\overline{F}(\theta) (1-\rho_{\theta})}}
{\left(\sum_{i=0}^{\infty}{\Phi_2\left(\Phi_1^i(\overline{F}(x+\theta))\right)}\right)}.
\end{eqnarray}
\end{Theorem}
\begin{proof}[\textsc{\textbf{Proof.}}]
From ({\ref{alpha_n_1}}) we have
\begin{eqnarray*}
&& \alpha'_n=q^n \alpha'_0+
\sum_{i=1}^{n-1}{q^i}+\frac{b(\theta)}{\overline{F}(\theta)}
\sum_{i=1}^{n-1}{\Phi_1^i(\overline{F}(x+\theta))}+\frac{b(\theta)}{\overline{F}(\theta)}
\overline{F}(x+\theta)+1,
\end{eqnarray*}
and then from (\ref{T_BPS_n_1}) and ({\ref{eq:Phi_1_const}}) it
follows, that
\begin{eqnarray*}
&& \overline{T}^{BPS}_n(\theta)=(m-{\overline{X_{\theta}^1}})
\left( q^n \alpha'_0+ \sum_{i=0}^{n-1}{q^i}
\right)+\frac{b(\theta)}{\overline{F}(\theta)} \left(
\Phi_2\left(\sum_{i=0}^{n-1}{\Phi_1^i(\overline{F}(x+\theta))}\right)\right).
\end{eqnarray*}
Using the facts (see (\ref{q_val})):
\begin{eqnarray*}
&& 1.\, q<\rho<1 \Longrightarrow q^n \rightarrow 0 \ \mbox{as}
\ n \rightarrow \infty, \\
&& 2.\, \sum_{i=0}^{\infty} {q^i} = \frac{1}{1-q}=
\frac{1-\rho_{\theta}}{1-\rho},
\end{eqnarray*}
we conclude that
\begin{eqnarray*}
{\overline{T}^{BPS}(\theta)}= \lim_{n\rightarrow \infty}
{\overline{T}^{BPS}_n(\theta)={{(m-{\overline{X_{
\theta}^1}})}}\left( \frac{1-\rho_{\theta}}{1-\rho}\right)
+{\frac{b(\theta)}{\overline{F}(\theta)}}\left(
{\sum_{i=0}^{\infty}{\Phi_2\left(\Phi_1^i(\overline{F}(x+\theta))\right)}}\right)}.
\end{eqnarray*}
Finally, using (\ref{eq:T_T_BPS}) we obtain (\ref{eq:T_oper}).
\end{proof}

The resulting formula (\ref{eq:T_oper}) is difficult to analyze
and does not have a clear analytic expression. Using this result
in the next subsection we find an approximation,which is also an
upper bound, of the expected sojourn time function in a more
explicit form.

\subsection{Upper bound for the expected sojourn time}

$\quad$ \mbox{} Let us start with auxiliary results.
\begin{Lemma}{\label{th_Phi_sigma_condit}}
For any function $\beta(x) \geq 0$ with
$\beta_j=\int_{0}^{\infty}{\beta(x)e^{-\mu_i x}dx}$,
\begin{equation*}
\textrm{if} \quad \frac{d (\beta_j \mu_j)}{d \mu_j} \geq 0, \quad
j=1,...,N  \quad  \textrm{ it follows, that } \quad {
\Phi_2\left({\Phi_{1}}( {\beta}(x))\right) } \leq q\Phi_2\left(
{\beta}(x) \right).
\end{equation*}
\end{Lemma}
\begin{proof}[\textsc{\textbf{Proof.}}]
See Appendix.
\end{proof}

\begin{Lemma}{\label{lm_ph_alpha}}
For the TLPS system with the hyper-exponential job size
distribution the following statement holds:
\begin{eqnarray}{\label{eq:Phi_leq_Phi}}
&& \Phi_2\left({\Phi_{1}}( {\alpha}'(x))\right)  \leq
q\Phi_2\left( {\alpha}'(x) \right).
\end{eqnarray}
\end{Lemma}
\begin{proof}[\textsc{\textbf{Proof.}}]
We define $\alpha'_j={\int_{0}^{\infty}} {{\alpha}'(x)}{e^{-\mu_j
x}}dx, \,\,\, j=1,...N$. As was shown in
{\cite{Osipova_rap-rech}}, $\alpha'(x)$ has the following
structure:
\begin{eqnarray*}
&& \alpha'(x)=a_0+\sum_{k}{a_k e^{-b_k x}}, \quad a_0\geq0,\,
a_k\geq0,\, b_k>0, \quad k=1,...,N.
\end{eqnarray*}
Then, we have that $\alpha'(x) \geq 0$ and
\begin{eqnarray*}
&& \alpha'_j=\frac{a_0}{\mu_j}+ \sum_{k}{\frac{a_k}{b_k+\mu_j}}, \quad j=1,...,N,\\
\Longrightarrow && \frac{d (\alpha'_j \mu_j)}{d \mu_j}=
\sum_{k}{\frac{a_k} {b_k+\mu_j}}-\sum_{k}{\frac{a_k \mu_j}
{(b_k+\mu_j)^2}}=\sum_{k}{\frac{a_k b_k} {(b_k+\mu_j)^2}} \geq 0,
\quad j=1,...,N,
\end{eqnarray*}
as \mbox{$a_k\geq0,\, b_k>0, \quad  k=1,...,N$}. So, then,
according to Lemma~{\ref{th_Phi_sigma_condit}} we have
({\ref{eq:Phi_leq_Phi}}).
\end{proof}

Let us state the following Theorem:
\begin{Theorem}{\label{teor:upper_bound}}
An upper bound for the expected sojourn time
${\overline{T}(\theta)}$ in the TLPS system with the
hyper-exponential job size distribution function with many phases
is given by $\overline{\Upsilon}(\theta)$:
\begin{eqnarray}\label{eq:tightbound}
&&{\overline{T}(\theta)} \leq {\overline{\Upsilon}(\theta)}=
\frac{{\overline{X_{ \theta}^1}}+\overline{W}( \theta
)\overline{F}( \theta )}{1-\rho_{ \theta
}}+{\frac{m-{\overline{X_{\theta}^1}}}{1-\rho}
+{\frac{b(\theta)}{\overline{F}(\theta)(1-\rho)}}}{\sum_{i,j}{\frac{\overline{F_{\theta}^i}\,
\overline{F_{\theta}^j}}{\mu_i+\mu_j}}}.
\end{eqnarray}
\end{Theorem}
\begin{proof}[\textsc{\textbf{Proof.}}]
According to the recursion (\ref{alpha_n_1}) we have for
$\alpha'_n(x)$ we approximate $\alpha'(x)$ with the function
${\widetilde{\alpha}'(x)}$, which satisfies the following
equation:
\begin{eqnarray*}
&&{\widetilde{\alpha}'(x)}= {\widetilde{\alpha}'(x)\Phi_1(1)
+\frac{b(\theta)}{\overline{F}(\theta)}\overline{F}(x+\theta)+1}.
\end{eqnarray*}
Then, according to (\ref{eq:Phi_1_const}) we can find the
analytical expression for ${\widetilde{\alpha}'(x)}$:
\begin{eqnarray*}
&& {\widetilde{\alpha}'(x)}= q{\widetilde{\alpha}'(x)
+\frac{b(\theta)}{\overline{F}(\theta)}
\overline{F}(x+\theta)+1}, \\
\Longrightarrow && {\widetilde{\alpha}'(x)} =
\frac{1}{1-q}\left(\frac{b(\theta)}{\overline{F}(\theta)}
\overline{F}(x+\theta)+1\right).
\end{eqnarray*}
We take \mbox{${\overline{\Upsilon}^{BPS}(\theta)}
=\Phi_2(\widetilde{\alpha}'(x))$} as an approximation for
${\overline{T}^{BPS}(\theta)=\Phi_2(\alpha'(x))}$. Then
\begin{eqnarray*}
{\overline{\Upsilon}^{BPS}(\theta)}=\Phi_2(\widetilde{\alpha}'(x))=
\frac{(m-\overline{X_{\theta}^1})}{1-q}+\frac{b(\theta)}{\overline{F}(\theta)}{\Phi_2(\overline{F}(x+\theta))}
=\frac{(m-\overline{X_{\theta}^1})}{1-q}+\frac{b(\theta)}{\overline{F}(\theta)}
{\sum_{i,j}{\frac{\overline{F_{\theta}^i}\,
\overline{F_{\theta}^j}}{\mu_i+\mu_j}}}.
\end{eqnarray*}
Let us prove, that
\begin{eqnarray*}
&& \overline{T}^{BPS}(\theta) \leq \overline{\Upsilon}^{BPS}
(\theta),
\end{eqnarray*}
or equivalently
\begin{eqnarray*}
&& \overline{T}^{BPS}(\theta) - \overline{\Upsilon}^{BPS}
(\theta)=\Phi_2(\alpha'(x)) - \Phi_2(\widetilde{\alpha}'(x)) \leq
0.
\end{eqnarray*}

Let us look at
\begin{eqnarray*}
&& \Phi_2(\alpha'(x)) - \Phi_2(\widetilde{\alpha}'(x)) = \\
&&=\Phi_2(\Phi_1(\alpha'(x))) +
\Phi_2\left(\frac{b(\theta)}{\overline{F}(\theta)}
\overline{F}(x+\theta)+1\right) -\left(q\Phi_2({
\widetilde{\alpha}'(x)}) +
\Phi_2\left(\frac{b(\theta)}{\overline{F}(\theta)} \overline{F}(x+\theta)+1\right)\right) \\
&&=\Phi_2(\Phi_1(\alpha'(x)))-q\Phi_2 ({ {\alpha}'(x)})+q
\left(\Phi_2({ {\alpha}'(x)})
-\Phi_2({ \widetilde{\alpha}'(x)})\right) \\
&&\Longrightarrow\\
 &&\Phi_2({ {\alpha}'(x)}) -\Phi_2({
\widetilde{\alpha}'(x)})=\frac{1}{1-q}\left(\Phi_2(\Phi_1(\alpha'(x)))-q\Phi_2
({ {\alpha}'(x)})\right).
\end{eqnarray*}

And from Lemma~\ref{lm_ph_alpha} and formula (\ref{eq:T_T_BPS}) we
conclude that ({\ref{eq:tightbound}}) is true.
\end{proof}

In this subsection we found the analytical expression of the upper
bound of the expected sojourn time in the case when the job size
distribution is a hyper-exponential function with many phases. In
the experimental results of the following subsection we show that
the obtained upper bound is also a close approximation. The
analytic expression of the upper bound which we obtained is more
clear and easier to analyze then the expression of the expected
sojourn time. It could be used in the future research on TLPS
model.

\subsection{Experimental results}

$\quad$ \mbox{} We calculate $\overline{T}(\theta)$ and
$\overline{\Upsilon}(\theta)$ for different numbers of phases $N$
of the job size distribution function. We take \mbox{$N=10, 100,
500, 1000$}. To calculate $\overline{T}(\theta)$ we find the
numerical solution of the system of linear equations
(\ref{Li_part1}) using the Gauss method. Then using the result of
Proposition {\ref{propos:linear_sys_solut}} we find
$\overline{T}(\theta)$. For $\overline{\Upsilon}(\theta)$ we use
equation (\ref{eq:tightbound}).

As was mentioned in Subsection~\ref{subsec:defin}, by using the
hyper-exponential distribution with many phases, one can
approximate a heavy-tailed distribution. In our numerical
experiments, we fix $\rho$, $m$, and select $p_i$ and $\mu_i$ in a
such a way, that by increasing the number of phases we let the
second moment $d$ (see ({\ref{eq:m_d}})) increase as well. Here we
take
\begin{eqnarray*}
&& \rho=10/11,\,\, \lambda=0.5,\,\, p_i=\frac{\nu}{i^{2.5}},\,\,
\mu_i=\frac{\eta}{i^{1.2}}, \quad i=1,...,N.
\end{eqnarray*}
In particular, we have
\begin{eqnarray*}
&& \sum_{i}{p_i}=1,\quad  \Longrightarrow\,\, \nu
=\frac{1}{\sum_{i}{i^{-2.5}}},\\
&&\sum_{i}{\frac{p_i}{\mu_i}}=m,\,\,\, \Longrightarrow\,\,
\eta={\frac{\nu}{m}}\,{\sum_{i}{i^{-1.3}}}.
\end{eqnarray*}

In Figure~{\ref{fg:phases_change}} one can see the plots of the
expected sojourn time and its upper bound as functions of $\theta$
when $N$ varies from 10 up to 1000.  In
Figure~{\ref{fg:_error_phases_change}} we plot the relative error
of the upper bound
\begin{eqnarray*}
&& \Delta(\theta)=\frac{\overline{\Upsilon}(\theta)-
\overline{T}(\theta)}{\overline{T}(\theta)},
\end{eqnarray*}
when $N$ varies from 10 up to 1000. As one can see, the upper
bound (\ref{eq:tightbound}) is very tight.

We find the maximum gain of the expected sojourn time of the TLPS
system with respect to the standard PS system. The gain is given
by \mbox{$g(\theta)={\frac{\overline{T}^{PS} -
\overline{T}(\theta)}{\overline{T}^{PS}}}$}. Here
$\overline{T}^{PS}$ is an expected sojourn time in the standard PS
system. Let us notice, that $\overline{T}^{PS}=\overline{T}(0)$.

The data and results are summarized in Table~{\ref{tab:table1}}.

\begin{table}[htb]
\centerline{\begin{tabular}{|r|r|r|r|r|r|} \hline
N & $\eta$ & d & $\theta_{opt}$ & $\max_{\theta}g(\theta)$ & $\max_{\theta}\Delta(\theta)$ \\
\hline \hline
10   & 0.95   & 7.20     &  5  & 32.98$\%$ & 0.0640 \\
100  & 1.26   & 32.28    &  12 & 45.75$\%$ & 0.0807 \\
500  & 1.40   & 113.31   &  21 & 49.26$\%$ & 0.0766 \\
1000 & 1.44   & 200.04   &  26 & 50.12$\%$ & 0.0743 \\
\hline
\end{tabular}}
\caption{Increasing the number of phases}\label{tab:table1}
\end{table}

With the increasing number of phases we observe that

1. the second moment $d$ increases;

2. the maximum gain $\max_{\theta} g(\theta)$ in expected sojourn
time in comparison with PS increases;

3. the relative error of the upper bound $\Delta(\theta)$ with the
expected sojourn time decreases after the number of phases becomes
sufficiently large;

4. the sensitivity of the system performance with respect to the
selection of the sub-optimal threshold value decreases.

Thus the TLPS system produces better and more robust performance
as the variance of the job size distribution increases.

 \begin{figure}[t]
   \begin{minipage}{0.48\linewidth}
        \centering {\epsfxsize=3.1 in \epsfbox{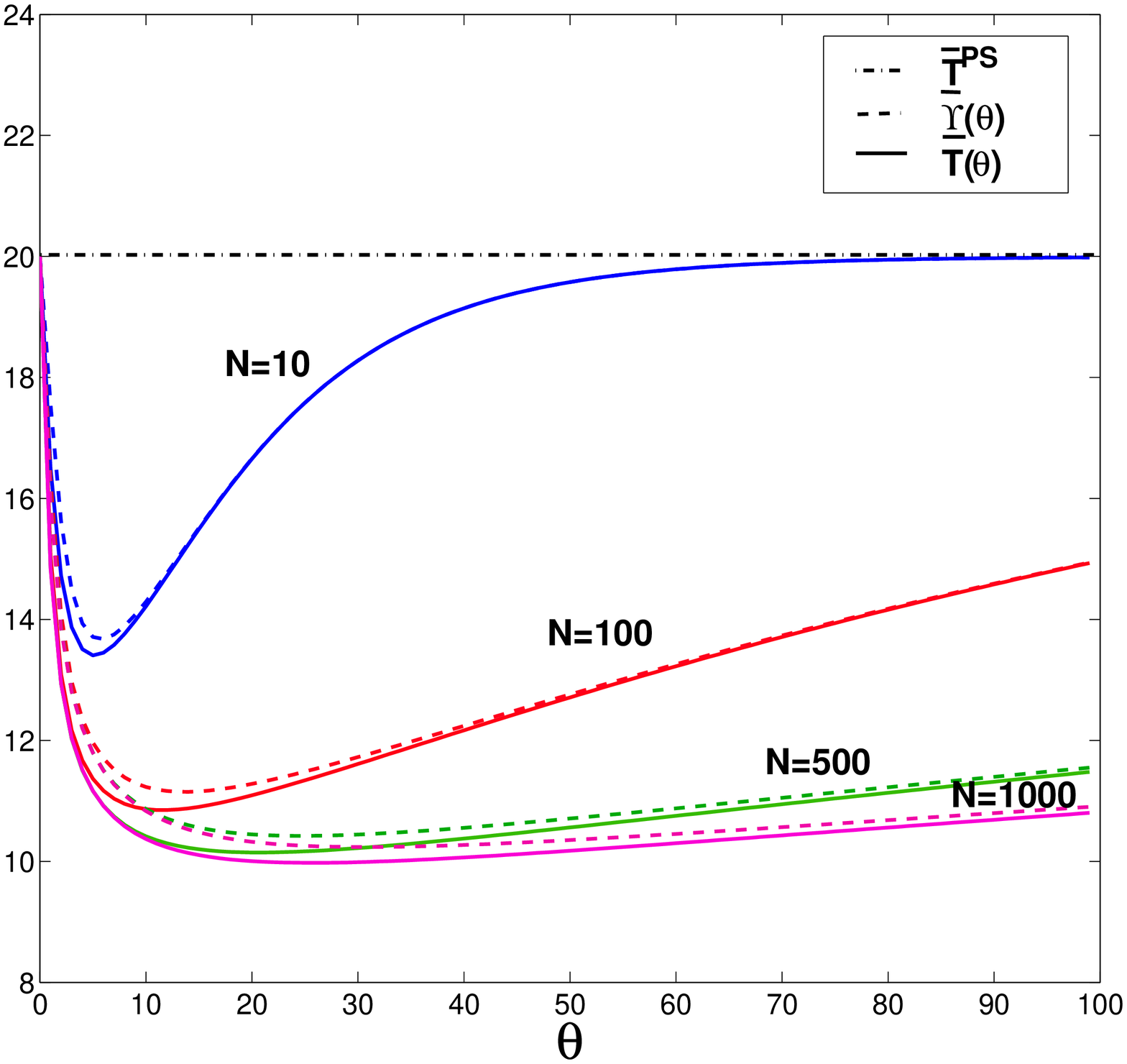}}
        \caption{The expected sojourn time $\overline{T}(\theta)$ and its upper bound $\overline{\Upsilon}(\theta))$ for $N=10$, $100$, $500$, $1000$}
        {\label{fg:phases_change}}
   \end{minipage}
   \begin{minipage}{0.48\linewidth}
        \centering {\epsfxsize=3.1 in \epsfbox{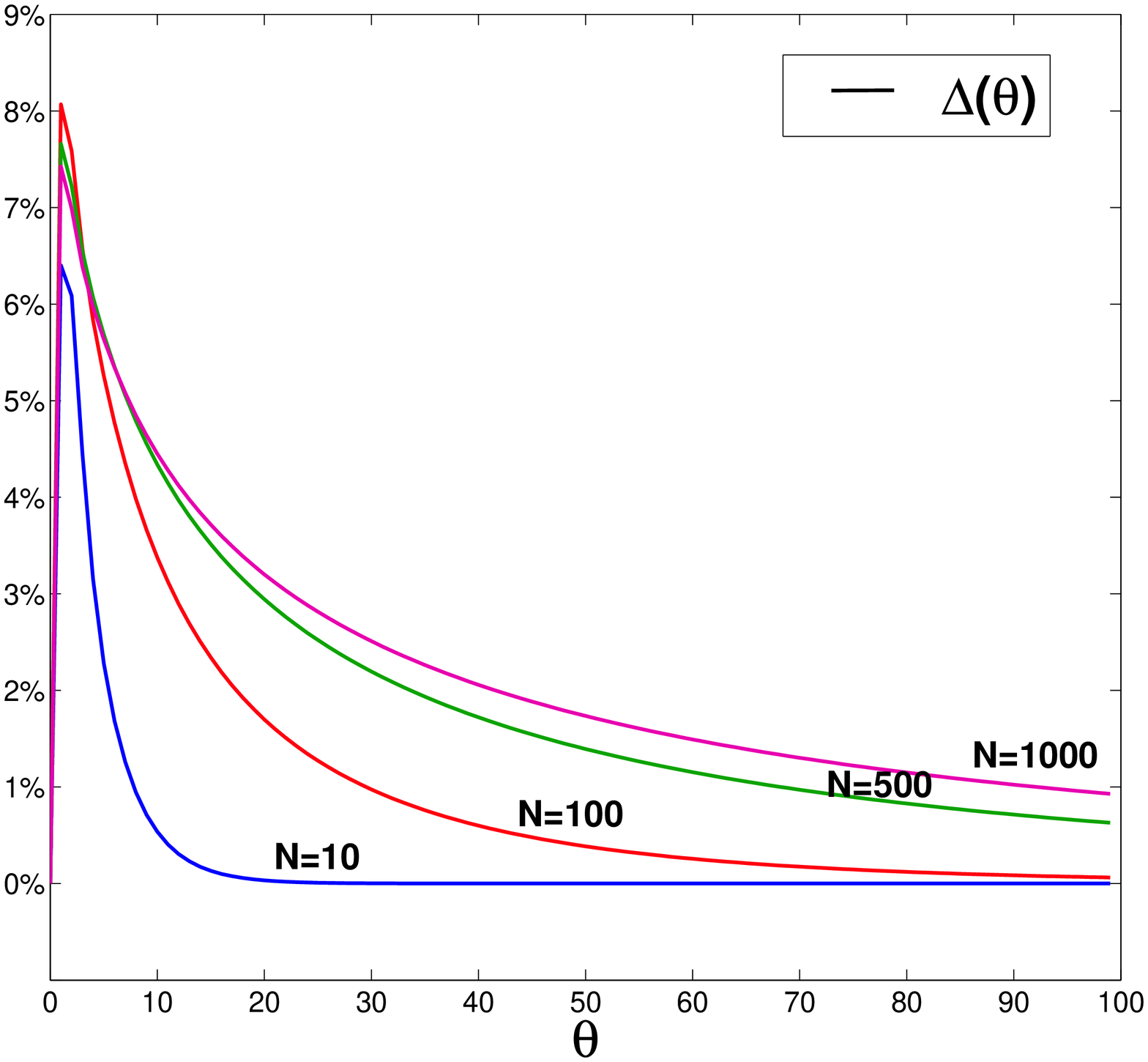}}
        \caption{The relative error\mbox{\,\,\,\quad\quad\quad\quad\quad} \mbox{$\Delta(\theta)=(\overline{T}(\theta)-\overline{\Upsilon}(\theta))/\overline{T}(\theta)$}  for $N=10$, $100$, $500$, $1000$}
        {\label{fg:_error_phases_change}}
   \end{minipage}
 \end{figure}

\newpage
\section{Conclusion}

$\quad$ \mbox{} We analyze the TLPS scheduling mechanism with the
hyper-exponential job size distribution function.

In Section {\ref{sec:twophases}} we analyze the system when the
job size distribution function has two phases and find the
analytical expressions of the expected conditional sojourn time
and the expected sojourn time of the TLPS system.


Connections in the Internet belong to two distinct classes: short
HTTP and P2P signaling connections and long downloads such as:
PDF, MP3, and so on. Thus, according to this observation, we
consider a special selection of the parameters of the job size
distribution function with two phases and find the approximation
of the optimal threshold, when the variance of the job size
distribution goes to infinity.

We show, that the approximated value of the threshold tends to the
optimal threshold, when the second moment of the distribution
function goes to infinity. We found that the gain of the TLPS
system compared to the standard PS system could reach 45\% when
the load of the system increases. Also the system is not too
sensitive to the selection of the optimal value of the threshold.

In Section {\ref{sec:manyphases}} we have studied the TLPS model
when the job size distribution is a hyper-exponential function
with many phases. We provide an expression of the expected
conditional sojourn time as a solution of the system of linear
equations. Also we apply the iteration method to find the
expression of the expected conditional sojourn time in the form of
operator series and using the obtained expression we provide an
upper bound for the expected sojourn time function. With the
experimental results we show that the upper bound is very tight
and could be used as an approximation of the expected sojourn time
function. We show numerically, that the relative error between the
upper bound and expected sojourn time function decreases when the
variation of the job size distribution function increases. The
obtained upper bound could be used to identify an approximation of
the optimal value of the optimal threshold for TLPS system when
the job size distribution is heavy-tailed.

We study the properties of the expected sojourn time function,
when the parameters of the job size distribution function are
selected in such a way, that it approximates a heavy-tailed
distribution as the number of phases of the job size distribution
increases. As the number of phases increases the gain of the TLPS
system compared with the standard PS system increases and the
sensitivity of the system with respect to the selection of the
optimal threshold decreases.

\newpage
\section{Appendix: Proof of Lemma~\ref{th_Phi_sigma_condit}}

$\quad$ \mbox{} Let us take any function \mbox{$\beta(x)
>0 $} and define $\beta_j={\int_{0}^{\infty}} {\beta(x)}{e^{-\mu_j
x}} dx, \quad j=1,...,N.$ Let us show for $\beta(x) \geq 0 $ that
if
\begin{eqnarray*}
&& \frac{d ({\beta}_j \mu_j)}{d \mu_j} \geq 0, \quad j=1,...,N,
\quad \textrm{ then it follows that } \quad
{\Phi_2\left({\Phi_{1}}( {\beta}(x))\right) } \leq q\Phi_2\left(
{\beta}(x) \right).
\end{eqnarray*}
As
\begin{eqnarray*}
&& {\int_{0}^{\infty}\!\! \int_{0}^{x}\!\! {\beta(y)}
{\overline{F} (x-y+\theta)}{\overline{F}(x+\theta)} dy dx =
\int_{0}^{\infty}\!\! \int_{0}^{\infty}\!\! {\beta(y)}
{\overline{F} (x_1+\theta)} {\overline{F}(x_1+y+\theta)} dx_1 dy}
\end{eqnarray*} and
\begin{eqnarray*}
\Phi_2(\Phi_{1}(\beta(x)))&=&{\gamma(\theta)\int_{0}^{\infty}\!\!
\int_{0}^{\infty}\beta(y) {\overline{F} (x+y+\theta)}{\overline{F}( x+\theta)}dy dx}\\
&& + {\gamma(\theta){\int_{0}^{\infty}\!\! \int_{0}^{x} \beta(y)
{\overline{F}( x-y+\theta)}{\overline{F}( x+\theta)}dy dx }},
\end{eqnarray*}
then
\begin{eqnarray*}
{\Phi_2(\Phi_{1}(\beta(x)))}&= & 2 {\gamma(\theta)
\int_{0}^{\infty} \!\!\int_{0}^{\infty} {\beta(x)} {\overline{F}
(x+\theta)} {\overline{F}(x+y+\theta)} dy dx} = \\
&=& 2{ \gamma(\theta) \int_{0}^{\infty} {\beta(x)} {\sum_{i,j}
{\frac{\overline{F_{\theta}^i}
\overline{F_{\theta}^j}}{\mu_i+\mu_j}}{e^{-\mu_j x}}} dx}= 2
\gamma(\theta)  {\sum_{i,j} {\frac{\overline{F_{\theta}^i}
\overline{F_{\theta}^j}}{\mu_i+\mu_j}} {\beta_j}}.
\end{eqnarray*}
Also for $\Phi_2\left( \beta(x) \right)$, taking into account that
\mbox{$q=\gamma(\theta)
\sum_i{\frac{\overline{F_{\theta}^i}}{\mu_i}}$}, we obtain
\begin{eqnarray*}
q\Phi_2\left( \beta(x) \right)&=& {\gamma(\theta){\sum_{i}
{\frac{\overline{F_{\theta}^i}}{\mu_i}}} {\sum_{j}
{\overline{F_{\theta}^j}} {\int_{0}^{\infty}} {\beta(x)}{e^{-\mu_j
x}}} dx} = \gamma(\theta) {\sum_{i,j}
{\frac{\overline{F_{\theta}^i}
\overline{F_{\theta}^j}}{\mu_i}}\beta_j}.
\end{eqnarray*}

Thus, a sufficient condition for the inequality \mbox{${
\Phi_2\left({\Phi_{1}}( \beta(x))\right) } \leq q\Phi_2\left(
\beta(x) \right) $} to be satisfied is that for every pair
\mbox{$i,j$}:
\begin{eqnarray*}
&&{{\frac{2}{\mu_i+\mu_j}} \beta_j+{\frac{2}{\mu_j+\mu_i}}
\beta_i}  \leq {\frac{1}{ \mu_i}} \beta_j +{\frac{1}{\mu_j}}
\beta_i  \quad \Longleftrightarrow \quad -(\beta_j \mu_j - \beta_i
\mu_i) (\mu_j-\mu_i) \leq 0.
\end{eqnarray*}

The inequality is indeed satisfied when $\beta_j \mu_j$ is an
increasing function of $\mu_j$. We conclude that
\mbox{${\Phi_2\left({\Phi_{1}}( {\beta}(x))\right) } \leq
q\Phi_2\left( {\beta}(x) \right)$}, which proves
Lemma~~\ref{th_Phi_sigma_condit}.



\newpage
\tableofcontents


\newpage
\begin{thebibliography}{1}

\bibitem{AA06}
S. Aalto and U. Ayesta, ``Mean delay analysis of multilevel
processor sharing disciplines'', In Proceedings of
IEEE INFOCOM 2006.

\bibitem{AAN_TLPS}
S. Aalto, U. Ayesta, and E.Nyberg, ``Two-level processor sharing
scheduling disciplines: mean delay analysis'', in Proceedings of ACM
SIGMETRICS/Performance 2004.

\bibitem{RuN2C}
K. Avrachenkov, U. Ayesta, P. Brown, and E. Nyberg,
``Differentiation between short and long TCP flows: Predictability
of the response time'', in Proceedings of IEEE INFOCOM 2004.

\bibitem{AAB_BPS_MLPS_QS}
K. Avrachenkov, U. Ayesta, P. Brown, ``Batch Arrival
Processor-Sharing with Application to Multi-Level
Processor-Sharing Scheduling'', Queuing ystems 50, pp.459-480,
2005.

\bibitem{BM06}
F. Baccelli and D.R. McDonald, ``A stochastic model for the rate of
non-persistent TCP flows'', in Proceedings of ValueTools 2006.

\bibitem{bansal}
N. Bansal, ``Analysis of the M/G/1 processor-sharing queue with bulk
arrivals'', {\it Operations Research Letters}, v.31, no.5, 2003,
pp.401-405.

\bibitem{HyperExp}
R. El Abdouni Khayari, R. Sadre, and B.R. Haverkort, ``Fitting
world-wide web request traces with the EM-algorithm'', {\it
Performance Evaluation}, v.52, no.2-3, 2003, pp.175-191.

\bibitem{feldmann_whitt}
A. Feldmann, W. Whitt, ''Fitting mixtures of exponentials to
long-tail distributions to analyze network performance models'',
{\it Performance Evaluation}, v.31, 1998, pp.245-258.

\bibitem{FM03}
H. Feng and V. Misra, ``Mixed scheduling disciplines for network
flows'', {\it ACM SIGMETRICS Performance Evaluation Review},
v.31(2), 2003, pp.36-39.

\bibitem{GM02}
L. Guo and L. Matta, ``Differentiated control of web traffic:
A numerical analisys, In SPIE ITCOM, Boston, 2002.

\bibitem{HB03}
M. Harchol-Balter, B. Schroeder, N. Bansal, and M. Agrawal,
``Size-based scheduling to improve web performance'',
{\it ACM Transactions on Computer Systems}, v.21, no.2,
2003, pp.207-233.

\bibitem{kleinrock}
L. Kleinrock, \textit{Queueing systems, vol. 2}, John Wiley and
Sons, 1976.

\bibitem{Kle_1}
Kleinrock, L., R. R. Muntz, and E. Rodemich, ``The
Processor-Sharing Queueing Model for Time-Shared Systems with Bulk
Arrivals,'' Networks Journal,1,1-13 (1971).

\bibitem{Osipova_rap-rech}
N. Osipova, ``Batch Processor Sharing with Hyper-Exponential
service time'', INRIA Research Report RR-6180, 2007.
Available at http://hal.inria.fr/inria-00144389


\bibitem{S68}
L.E. Schrage, ``A proof of the optimality of the shortest
remaining processing time discipline'', {\it Operations Research},
v.16, 1968, pp. 687-690.

\end{thebibliography}
\end{document}